# Local Detection of Enhanced Hot Electron Scattering in InSb/CdTe Heterostructure Interface

Xiaoxiao Ma[1, 2, #], Zhenghang Zhi[3, #], Weijie Deng[1, 2], Tianxin Li[2, a)], Qianchun Weng[2, a)], Xufeng Kou[3, a)], Wei Lu[1, 2, a)]

[1]School of Physical Science and Technology, ShanghaiTech University, Shanghai, 201210, China

[2]State Key Laboratory of Infrared Physics, Shanghai Institute of Technical Physics, Chinese Academy of Sciences, Shanghai 200083, China

[3]ShanghaiTech Laboratory for Topological Physics, ShanghaiTech University, Shanghai 200031, China.

[#] These authors contributed equally to this work

[a)] Authors to whom correspondence should be addressed. Electronic addresses:

qcweng@mail.sitp.ac.cn, txli@mail.sitp.ac.cn, kouxf@shanghaitech.edu.cn and luwei@mail.sitp.ac.cn

**Abstract**

    The InSb/CdTe heterojunction structure, characterized by low effective mass and high electron mobility, exhibits interfacial energy band bending, leading to the Rashba spin-orbit coupling effect and nonreciprocal transport, which makes its suitable for spintronic devices with broad applications in logic and storage fields. However, the complex heterojunction interfaces of InSb/CdTe, composed of group III-V and group II-VI semiconductors, are prone to interdiffusion. Therefore, characterization and study of the interfacial properties of InSb/CdTe heterojunctions are crucial for the growth improvement of the InSb/CdTe material system as well as its application in the field of spintronics. In this study, a novel scanning probe microscope, called a scanning noise microscope, was applied to visualize hot electron scattering in InSb/CdTe nano-devices. The results demonstrated that the near-field signal originates from the Coulomb scattering of charged ions on electrons at the interface of the embedded layer heterojunction. This real-space, nondestructive characterization of the heterojunction interface properties offers a new tool for enhancing the performance of heterojunctions.

Semiconductor heterostructures offer immense potential for a wide range of applications in high-speed and optoelectronic devices. As early as the 1930s, Zhores I. Alferov et al. discovered that the electronic properties of semiconductor heterostructures are significantly influenced by the interface between the two materials[1]. A heterojunction is formed by two different materials, and one key characteristic of this system is the disruption of spatial inversion symmetry at the interface. This is achieved through the establishment of a built-in potential, resulting in the interface Rashba spin-orbit coupling (SOC) effect. Consequently, it furnishes a potent avenue for the electrical manipulation of spin states, which is of great importance for advancing next-generation high-performance spintronic devices.

Understanding and modulating interfacial properties are crucial for optimizing the efficacy of electronic and optoelectronic devices based on the heterojucntion structures[2,3]. InSb/CdTe heterostructures have attracted siginificant interest because of their diverse properties. Previous studies have shown that InSb/CdTe heterostructures exhibit a type-I ladder energy band structure[6,7]. The substantial internal polar field at the interface can induce energy band reversal, resulting in the manifestation of topologically nontrivial edge states characterized by pronounced Rashba-type spin splitting[8]. This interfacial Rashba effect induce the InSb/CdTe system with robust spin coupling capabilities, facilitating efficient conversion of injected charge currents into spin-polarized currents[9,10]. Furthermore, the spin-polarized currents generated by InSb/CdTe can exert a torque that modulates the magnetization direction within the ferromagnetic layer[11]. The spin currents generated by this system can influence the magnetization direction within the ferromagnetic layer[11]. Moreover, the Rashba spin-orbit coupling (SOC) at this interface introduces non-reciprocal magnetoelectric transport properties. Such characteristics render InSb/CdTe advantageous for diverse applications, including SOT-MRAM[12,13], dual-ended spin rectifiers[14], magnetic sensing[15], and spin field-effect devices[16,17]. The potent SOC effect in the InSb/CdTe material system arises from the energy band bending at the heterojunction interface. However, previous studies have shown that interdiffusion between InSb and CdTe at the heterojunction interface leads to the formation of interfacial charged ions, including negatively charged InSb surface and positively charged CdTe surface at the interface[8,18]. These charged ions are expected to change the transport properties of electrons at the interface, however, still lacks of detailed experimental studies, which is important for further enhancement of device performance

Nevertheless, probing local transport properties of electrons is challenging due to the absence of efficient experimental tool. Recently, a novel scanning probe microscope, called a scanning noise microscope (SNoiM) has been developed to directly visualize nonequilibrium charge transport and scattering in real-space[19]. SNoiM is sensitive to local charge carrier dynamics because it detects the fluctuating electromagnetic (EM) evanescent fields that is generated by nanoscale current fluctuations[20]. For instance, hot electrons in GaAs/AlGaAs quantum well devices experience strong electron-impurity/electron-phonon scatterings, generating intense fluctuating EM evanescent fields at the sample surface and detected by SNoiM. With real-space mapping, nonlocal[19] and quasiadiabatic[21] transport have been revealed. In this study, SNoiM is applied to investigate the detailed local electron transport in InSb/CdTe nanodevices, our results show that the detected fluctuating EM evanescent fields are mainly generated by two-dimensional electron gas (2DEG) in InSb/CdTe interface that is buried below surface, rather than conduction electrons at the surface (InSb bulk layer). Evanescent fields generated by conventional electron-impurity/electron-phonon scattering in InSb bulk are too weak to be detected, while 2DEG at the interface experiences strong Coulomb scattering from positive/negative ions due to interdiffusion and directly probed by SNoiM (as schematically shown in Figure 1(a)).

The InSb/CdTe heterojunction structure is grown along the GaAs (111) crystal direction by molecular beam epitaxy[22]. The energy-dispersive X-ray spectroscopy (EDS) diagram of the interface is shown in Fig. 1(b). The exceptional lattice matching[23] between InSb and CdTe enables the growth of strain-free, defect-free epitaxy layers. Both CdTe and InSb materials are undoped, however, due to small band gap of InSb (only ~ 0.17 eV), a significant number of conduction electrons are thermally excited at room temperature, resulting in highly conductive surface. As a result, two conducting channels exist in present InSb/CdTe devices: bulk InSb near surface and 2DEG at the InSb/CdTe interface. SNoiM is a surface technique and should be more sensitive to highly conductive InSb surface, however, as will be shown later, the detected fluctuating EM evanescent fields are mainly generated by 2DEG at the interfaced due to strong scattering from interfacial charged ions. The detailed sample structure is shown in Figure 2(a), which is fabricated into nanoscale constriction channel by electron beam lithography and dry etching. Ohmic contacts are simply prepared by evaporation of Ti/Au (10 nm/200 nm). Fig. 2(d) shows a scanning electron microscopy (SEM) image of the central nano-constricted region.

SNoiM can be regarded as the near-field version of infrared radiation thermometer[24]. Without approaching the tip to scatter the EM evanescent fields, it sensitively detects the far-field (FF) radiations from the sample surface. We first monitor the FF signal of the channel against different bias voltages ($V_{SD}$) and no obvious heating is observed as shown in Fig. 2(c), considering the spatial resolution ~ 20 μm and temperature sensitivity about ~ K. This is also clearly shown in the two dimensional(2D) FF image under $V_{SD}$= 3 V (Fig. 2(b)) and the temperature distribution is uniform. By introducing a sharp metal tip to scatter the fluctuating EM evanescent fields at the sample surface, clear 2D near-field images are obtained under different $V_{SD}$ (Fig. 2(e)) and a distinct local hot spot is observed in the nano-constriction. This is due to the generation of hot electrons by high electric-fields (~ 10 kV/cm) in the constriction. Nonlocal transport feature similar to that observed in high-field (~ 100 kV/cm) n-GaAs nanodevice[19] is not distinct in present samples due to much shorter hot electron relaxation length (< 50 nm), which is close or even below the spatial resolution of SNoiM (50-100 nm). By placing the tip above the nano-constriction center, we further monitor the near-field signal against $V_{SD}$, as shown in Fig. 2(c). The near-field signal is roughly proportional to the current (or voltage), which is consistent with the conventional theory of hot electron shot noise, $\langle S_{shot} \propto 2e|I|\rangle$. The signal decays rapidly when removing the tip from the sample surface and vanishes at the heights above 40 nm (Fig. 2(f)), indicating the evanescent nature of the detected signal.

As mentioned earlier, two parallel conducting channels exist in present sample (schematically shown in Figure 3(a)): bulk conduction in InSb and 2DEG conduction in InSb/CdTe interface. We found that the near-field signals can be only detected on those samples with thin InSb (< 100 nm), and vanishes if InSb/CdTe interface is buried far from the sample surface (results on a 400 nm-thick InSb sample is shown in Fig. 3(d)), indicating that the signal originates from the buried interface. In thicker InSb/CdTe devices, the electronic random motion of the 2D electron gas at the heterojunction interface cannot be detected by the tip because the heterojunction interface is buried deep below 400 nm on the sample surface, since the near-field signal decays rapidly within 100 nm as shown in Fig. 2(h), and the thicker InSb/CdTe devices only indicate the electronic information of the intrinsic InSb conductive channel. The experimental results show that the signals of the thicker InSb/CdTe devices are very weak and almost negligible, indicating that the evanescent waves generated by the random motion of electrons on the surface of intrinsic InSb are not detected, and that the scattering of phonons within intrinsic InSb is much weaker than that from the impurities at the interface. Therefore, it can be inferred that the infrared evanescent wave induced by phonon scattering in the two-dimensional electron gas of InSb/CdTe heterojunction is also difficult to be detected., it can be speculated that the strong near-field signal of the InSb/CdTe heterojunction 2D electron gas device does not come from the phonon scattering, but may come from the electronic random motion formed by the intense scattering of charged ions at the interface of the heterojunction to the 2D electron gas, which generates a very strong near-field signal, as shown schematically in Fig. 3(b).

This work investigates the terahertz noise behavior of the electronic motion of the device driven by an electric field, through infrared near-field optical tests of InSb/CdTe heterojunction nanochannels. In the experiments, the heterojunction devices with thinner layers of InSb show strong near-field signals, while the heterojunction devices with thicker layers of InSb show very weak and almost undetectable near-field signals generated by the electron motions on the surface, proving that the near-field signals mainly originate from the strong Coulomb scattering of the charged ions formed by the interfacial interdiffusion on the two-dimensional electron gas. This nondestructive and distinctive characterization method to directly study the heterojunction interface plays an important role in the study of heterojunction materials. This experiment is of great significance for us to clearly understand the nature of InSb/CdTe heterojunction interfaces, the scattering mechanism and to improve the performance of spintronics devices in the future.

## Data Statement:

The data that support the findings of this study are available from the corresponding author upon reasonable request.

# Figure captions

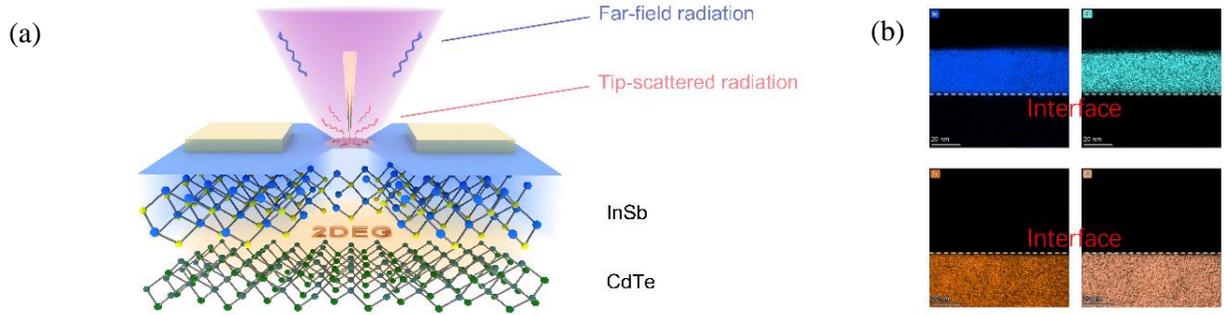

FIG. 1. (a) The experimental setup for the Scanning Noise Microscope (SNoiM) to detect evanescent waves in InSb/CdTe devices. (b)The energy dispersive X-ray spectroscopy (EDS) mapping of the InSb/CdTe interface.

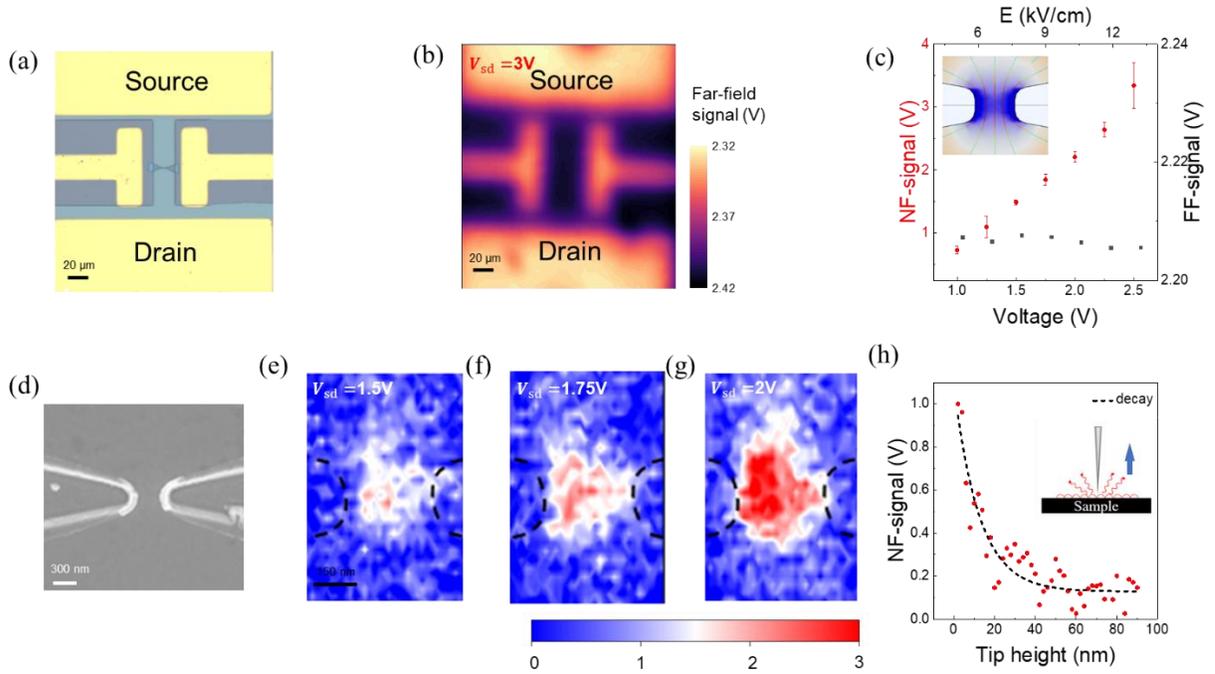

FIG. 2. (a) Optical image of the device (b) Far-field 2D image of the device (c) Near-field signal and far-field signal in the strong electric field region at the center of the channel as a function of bias voltage (electric field), Inset is the field strength in the center of the channel simulated by Comsol (d) SEM image of the channel (e) Near-field 2D image at different bias voltages (f) Decay line.

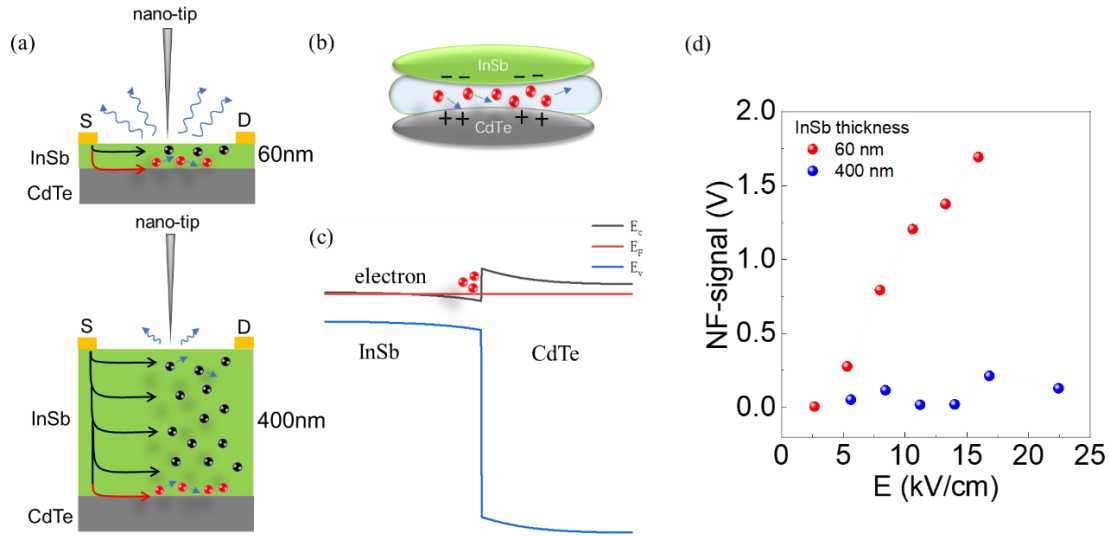

FIG. 3. (a) The current distribution model of the InSb/CdTe device, (b) Schematic of the strong Coulomb scattering of 2D electron gas by charged ions at the InSb/CdTe heterojunction interface. (c) The InSb/CdTe energy band structure, and (d) Near-field Signal as functions of Electric Field for 60 nm and 400 nm InSb/CdTe Devices. The near-field signal of the 60 nm thick InSb/CdTe device increases linearly with voltage (electric field), and the signal value tested in the 400 nm thick InSb/CdTe device is very weak, which can be considered as system background noise and neglected. Undetectable near-field signal is tested in the 400 nm thick InSb/CdTe device.